\documentclass[jgrga]{agutex}
\usepackage{amsmath} 
\usepackage{amssymb} 
\usepackage[]{graphicx}
\authorrunninghead{Komar et al.}
\titlerunninghead{Tracing magnetic separators}
\usepackage{verbatim}

\begin{document}
\hskip 8.5cm {\it Draft - Accepted July 27, 2013} \vskip .125cm

\title{Tracing magnetic separators and their dependence on IMF clock
  angle in global magnetospheric simulations}

\authors{C. M. Komar,\altaffilmark{1} P. A. Cassak,\altaffilmark{1}
  J. C. Dorelli,\altaffilmark{2} A. Glocer\altaffilmark{2} and
  M. M. Kuznetsova\altaffilmark{2}}
\altaffiltext{1}{Department of Physics and Astronomy, West Virginia University,
  Morgantown, West Virginia, USA}
\altaffiltext{2}{NASA Goddard Space Flight Center, Greenbelt,
  Maryland, USA}

\authoraddr{P. A. Cassak, Department of Physics and Astronomy, White Hall, Box 6315,
  West Virginia University, Morgantown, WV
  26506. (Paul.Cassak@mail.wvu.edu)}
\authoraddr{J. C. Dorelli, NASA Goddard Space Flight Center}
\authoraddr{A. Glocer, NASA Goddard Space Flight Center}
\authoraddr{C. M. Komar, Department of Physics and Astronomy, White Hall, Box 6315,
  West Virginia University, Morgantown, WV
  26506. (ckomar@mix.wvu.edu)}
\authoraddr{M. M. Kuznetsova, NASA Goddard Space Flight Center}

\begin{abstract}

  A new, efficient, and highly accurate method for tracing magnetic
  separators in global magnetospheric simulations with arbitrary clock
  angle is presented. The technique is to begin at a magnetic null and
  iteratively march along the separator by finding where four magnetic
  topologies meet on a spherical surface. The technique is verified
  using exact solutions for separators resulting from an analytic
  magnetic field model that superposes dipolar and uniform magnetic
  fields. Global resistive magnetohydrodynamic simulations are
  performed using the three-dimensional BATS-R-US code with a uniform
  resistivity, in eight distinct simulations with interplanetary
  magnetic field (IMF) clock angles ranging from 0 (parallel) to 180
  degrees (anti-parallel). Magnetic nulls and separators are found in
  the simulations, and it is shown that separators traced here are
  accurate for any clock angle, unlike the last closed field line on
  the sun-Earth line that fails for southward IMF. Trends in magnetic
  null locations and the structure of magnetic separators as a
  function of clock angle are presented and compared with those from
  the analytic field model. There are many qualitative similarities
  between the two models, but quantitative differences are also noted.
  Dependence on solar wind density is briefly investigated.

\end{abstract}

\begin{article}

\section{Introduction}
\label{section::Introduction}
Many important dynamic processes in the Earth's magnetosphere are
known or thought to be driven by magnetic reconnection, from solar
wind-magnetosphere coupling~\citep{gonzalez1990,borovsky2008b}, to
magnetospheric convection~\citep{dungey1961}, to substorm phenomena
(\citet{angelopoulos2008}, and references therein). Determining where
reconnection happens as a function of solar wind conditions is
critical for predicting its efficiency and for informing satellites,
such as NASA's upcoming Magnetospheric Multiscale (MMS) mission
\citep{burch2009,moore2013}, where to expect reconnection events to
occur.

In the classical model by~\citet{dungey1961,dungey1963}, reconnection
occurs at the subsolar point for southward interplanetary magnetic
field (IMF) and near the polar cusps for northward IMF. However, much
less is known about where reconnection occurs for arbitrary IMF clock
angles $\theta_\text{IMF}.$ For arbitrary $\theta_\text{IMF}$,
reconnection is likely to occur along the magnetic separator, the
magnetic field line that connects magnetic nulls (where the magnetic
field strength $\left|\mathbf{B}\right|=0$) and separates regions of
differing magnetic
topologies~\citep{cowley1973,siscoe1987,lau1990,siscoe2001}.  The
topology of a magnetic field line is determined by where it maps
relative to Earth: closed field lines map to the Earth in both
directions, open field lines do not map to Earth in either direction,
and half-open field lines only map to Earth in one direction. The
magnetic separator marks where magnetic topology changes, and the line
integral of the parallel electric field $E_\Vert$ along the separator
has been shown to be the global reconnection rate~\citep{siscoe2001}.

A number of methods have been developed to locate magnetic
separators. The eigenvectors of the $3\times3$ $\nabla\mathbf{B}$
tensor at a magnetic null determine the local magnetic field
geometry~\citep{greene1988, lau1990, parnell1996}. Other methods
determine the magnetic separator globally. One method convects solar
wind field lines Earthward to determine the separatrix
surfaces~\citep{dorelli2007,ouellette2010}. The last closed field line
along the sun-Earth line has also been used as an approximation of the
separator, as this field line closely approaches the magnetic nulls in
global magnetospheric simulations (\citet{dorelli2007} used this
method for northward IMF;~\citet{hu2009} determined the separator for
northward and southward IMF). The separator has also been located by
finding where different magnetic topologies
meet~\citep{laitinen2006,laitinen2007,dorelli2008,dorelli2009}.
\citet{haynes2010} developed an iterative technique to map the
separator using rings along the separatrix eigenvectors of the
$\nabla\mathbf{B}$ tensor. Another study~\citep{moore2008} inferred
reconnection geometries from deflections in streamlines at the
magnetopause.  A few of these studies have investigated separators as
a function of IMF clock
angle~\citep{laitinen2007,hu2009,ouellette2010}.

Magnetic nulls and separators arise in other contexts, as
well. \citet{longcope1996} mapped magnetic separators in simulations
of the solar corona by interpolating between the calculated separatrix
surfaces at both nulls.  \citet{close2004} traced magnetic separators
by locating changes in magnetic connectivity near magnetic nulls
resulting from magnetic potential fields in solar atmospheric
simulations. There has been observational evidence of magnetic nulls
and separators in Earth's magnetotail~\citep{xiao2006,xiao2007}.
Detecting these structures observationally is difficult because
multiple spacecraft are needed.

In this paper, we present a simple, efficient and accurate algorithm
to map magnetic separators in global magnetospheric simulations at the
dayside magnetopause for arbitrary IMF clock angle. We verify the
technique using exact solutions for an analytic model involving the
superposition of uniform and dipolar magnetic fields. Then, we trace
separators in global magnetohydrodynamic (MHD) simulations for various
IMF clock angles, and show that the last closed field line does not
map the separators for southward IMF. We discuss trends in magnetic
null locations and magnetic separators, making comparisons to the
analytical field model.

The layout of the paper is as follows. In
Sec.~\ref{section::Finding_Separators}, we present and verify the new
method for tracing magnetic separators. In
Sec.~\ref{section::Simulation_Study}, we describe the global
magnetospheric simulations, including a careful discussion about
numerical versus explicit dissipation. The results of finding null
locations and tracing separators for different IMF clock angles are
discussed in Sec.~\ref{section::Results}. The results are summarized
and potential applications are discussed in
Sec.~\ref{section::Conclusions}. The properties of magnetic
reconnection at the separators is outside the scope of the present
study.

\section{A Technique for Finding Separators}
\label{section::Finding_Separators}

\subsection{Technique Description}
\label{subsection::Technique_Description}

\begin{figure}
\centering
\noindent\includegraphics[width=20pc]{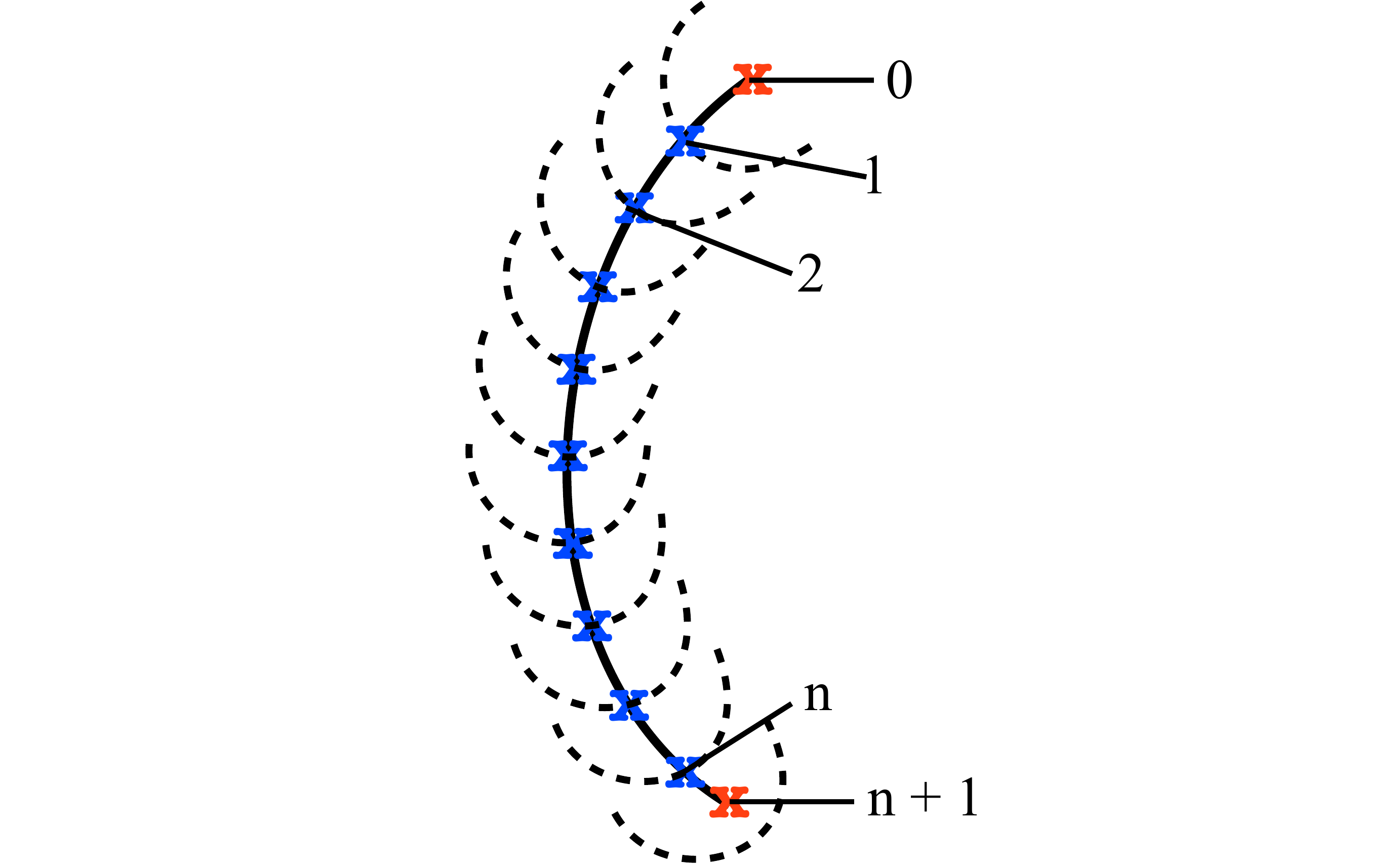}
\caption{Schematic diagram of the separator tracing algorithm.}
\label{fig::2D_Analog}
\end{figure}

The separator tracing algorithm exploits the fact that magnetic nulls
are the endpoints of magnetic separators. A schematic diagram of the
tracing process is shown in Fig.~\ref{fig::2D_Analog}.  The two
magnetic nulls are found using existing techniques, represented by
(red) X's at the endpoints. A hemisphere, represented by a dashed
semicircle, is centered around the northern null, labeled 0 (the
choice of starting null is arbitrary).  At many points on the
hemisphere's surface, the magnetic field is traced in both directions
to determine its topology (open, closed, or half-open). The point at
which all topologies meet is where the separator intersects the
hemisphere; we mark this intersection with a (blue) x and label it
location 1. A new hemisphere is centered at location 1 and the
separator's intersection with this new hemisphere is determined
similarly. This new intersection is marked by another (blue) x,
labeled as location 2.  This process is repeated until the southern
null, labeled $\text{n}+1$, is inside a hemisphere. The separator is
mapped by connecting the nulls to the individual separator locations
in order (0 through $\text{n}+1$), sketched as the solid (black) line.  (Note, an alternate algorithm would be to initially find a single null and perform the above iterative procedure while checking inside each hemisphere for another null at each step, but we do not pursue this further.)

We now provide a more detailed description of the steps in this
technique. To locate the magnetic nulls, we use the method described
by \citet{haynes2007b}. Since global minima of the magnetic field
magnitude $\left|\mathbf{B}\right|$ do not faithfully locate magnetic
nulls, every grid cell where all three components of the magnetic
field change sign is flagged. The field is then linearized within
these flagged cells and a Newton-Raphson iteration algorithm is used
to locate magnetic nulls at sub-grid resolution.

In most of our global magnetospheric MHD simulations, to be described
in Sec.~\ref{section::Simulation_Study}, this method returns a single
northern null and southern null. In one, multiple nulls are
identified, though the nulls in each hemisphere are within 0.25 Earth
radii $\left(\text{R}_\text{E}\right)$ of each other. To choose a
null, we select the location with the lowest $\left|\mathbf{B}\right|$
in one of the hemispheres. The simulations have a high degree of
symmetry, so the null in the opposite hemisphere is chosen to be
closest to the null's reflection. [That is, for a null located at
$\left(x,\,y,\,z\right)$, the reflected null is near
$\left(x,\,-y,\,-z\right)$]. The null identification is verified by
plotting field lines in the vicinity of the chosen nulls; magnetic
fields with different topologies converge in these regions, as
expected.

Having identified the nulls, we proceed to map the separator. Each
hemisphere in the iteration has a fixed radius $R_\text{HS}$
(hemispheres are used to automatically prevent retracing in the
opposite direction). Points on the hemisphere's surface are mapped by
a set of angular coordinates $\left(\phi,\,\lambda\right)$, where
$\phi$ is the longitude measured from the $+x$-axis in Geocentric
Solar Ecliptic (GSE) coordinates and $\lambda$ is the latitude (in GSE
coordinates, $x$ is sunward, $y$ is duskward, and $z$ is
northward). In spherical coordinates, $\lambda=90^\circ-\theta$, where
$\theta$ is the polar angle measured from the $+z$-axis in GSE.

\begin{figure}
\centering
\noindent\includegraphics[width=20pc]{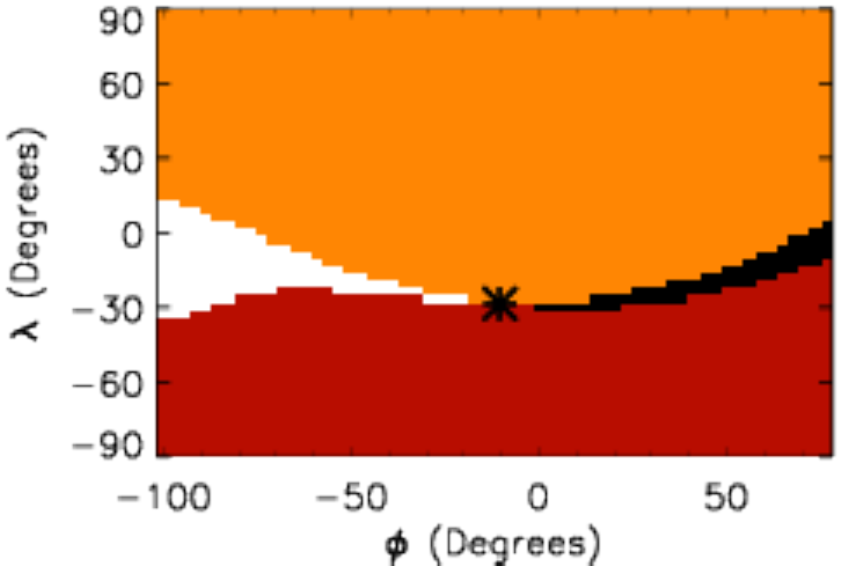}
\caption{Sample topology map for $\theta_\textsc{IMF}=30^\circ$ for a
  hemisphere centered at $\bold{r}=(3.16,\,~1.87,\,~8.01)~\
  \text{R}_\text{E}$, with radius of $1~\ \text{R}_\text{E}$. Colors
  denote magnetic topology: red are closed terrestrial fields, orange
  are open solar wind fields, black and white are half-closed fields
  that terminate at the south and north poles of Earth,
  respectively. The black asterisk marks the approximate separator
  location.}
\label{fig::Sample_Topology_Map}
\end{figure}

We use hemispheres with a radius of 1~R$_\text{E}$, and the surface of
each hemisphere is discretized into a $N_\phi\times N_\lambda$ grid
and the topology of the magnetic field is determined at each grid
point (we use a $61\times61$ grid). To calculate magnetic topology, we
use the Kameleon software package developed at NASA's Community
Coordinated Modeling Center (CCMC). We perform a bi-directional trace
of the magnetic field at each point on the surface, using the field's
footpoints to determine its topology. A field line is closed if both
footpoints are within 5~$\text{R}_\text{E}$ of the origin and is open
if both do not. Half-closed field lines have one footpoint within
5~$\text{R}_\text{E}$ of the origin. If the footpoint close to the
origin has a negative $z$-coordinate, then the field line is a
southern half-closed field line. Conversely, a northern half-closed
field line has the connecting footpoint with a positive
$z$-coordinate. Each point on the hemisphere is coded by its
topology. An example is in Fig.~\ref{fig::Sample_Topology_Map}; closed
magnetic fields are colored red, open magnetic fields are orange,
southern half-closed fields are black, and northern half-closed fields
are white.

To identify where the four topologies meet, interpolation is usually
necessary. The separator lies in between the northern and southern
half-closed regions. We start by searching through the topological map
for the locations where these regions are closest. In
Fig.~\ref{fig::Sample_Topology_Map}, the two closest points are at
$\left(0^\circ,\,~-30^\circ\right)$ and
$\left(-20^\circ,\,~-28^\circ\right)$. We find the topology of the
field line through the midpoint of the line connecting these
points. Then, we march out along the line perpendicular to this line
until the topology changes. The separator location is defined as the
average of the two points with differing topologies.
Figure~\ref{fig::Sample_Topology_Map} displays a (black) asterisk at
its approximate separator location which reasonably estimates where
the four topologies meet.

The separator location is used as the center of the subsequent
hemisphere. If the separator intersects the $k$-th hemisphere at
longitude and latitude $\left(\phi_k,\,~\lambda_k\right)$, the
coordinates of the next hemisphere's center $\mathbf{r}_{k+1}$ are
\begin{equation}
  \mathbf{r}_{k+1}=\mathbf{r}_{k}+\mathbf{r} \left(R_\text{HS},\,~\phi_{k},\,~\lambda_{k}\right),
\label{eqn::Sphere_Center}
\end{equation}
where $\mathbf{r}_{k}$ is the center of the $k$-th hemisphere, and
$\mathbf{r} \left(R_\text{HS},\phi_{k},\lambda_{k}\right)$ is the
separator's location on the $k$-th hemisphere in spherical coordinates
relative to $\mathbf{r}_k$. A range of
$\left[\phi_k-90^\circ,\,~\phi_k+90^\circ\right]$ is used as the
longitude of $k+1$-st hemisphere; this explains why the horizontal
axis in Fig.~\ref{fig::Sample_Topology_Map} is not centered around
$\phi=0^\circ$. An arbitrary degree of accuracy can be obtained by
decreasing the radius $R_\text{HS}$ and/or increasing the number of
grid points $N_\phi$, $N_\lambda$ on the hemispheres.

The method described here has some similarities to the four field
junction method by~\citet{laitinen2006}. This method calculates the
magnetic topology at every point on a Cartesian grid near regions
where the separator is thought to exist. The separator is approximated
by locations where all four topologies are within 3 grid cells of each
other, resulting in a ribbon-like structure at the dayside
magnetopause. Our method does not require a priori knowledge of the
separator's location, as it starts from the nulls and traces the
complete separator. It is also computationally inexpensive, since the
magnetic topology is calculated on a number of surfaces rather than a
volume at the dayside magnetopause.

\subsection{Verification with Vacuum Superposition}
\label{subsection::Technique_Verification}

To test the technique in Sec.~\ref{subsection::Technique_Description},
we use a simple magnetic field model with analytic solutions for the
nulls and separators. We superpose a dipolar magnetic field
$\mathbf{B}_\text{D}$ with a uniform background magnetic field
$\mathbf{B}_\text{IMF}$. The vacuum superposition magnetic field
$\mathbf{B}_\text{VS}$ is given by
\begin{equation}
\mathbf{B}_\text{VS}\left(\mathbf{r}\right)=\mathbf{B}_\text{D}\left(\mathbf{r}\right)+\mathbf{B}_\text{IMF},
\label{eqn::Vacuum_Superposition}
\end{equation}
where
\begin{equation}
\mathbf{B}_\text{D}\left(\mathbf{r}\right)=\frac{3\left(\mathbf{M}\cdot\mathbf{\hat{r}}\right)\mathbf{\hat{r}}-\mathbf{M}}{r^3},
\label{eqn::Dipole}
\end{equation}
$\mathbf{M}$ is Earth's magnetic dipole moment, and $\mathbf{r}$ is
the position vector. The IMF in GSE coordinates is
\begin{equation}
\mathbf{B}_\textsc{IMF}=B_\textsc{IMF}\left(\sin{\theta_\textsc{IMF}}\ ~\mathbf{\hat{y}}+\cos{\theta_\textsc{IMF}}\ ~\mathbf{\hat{z}}\right)
\label{eqn::IMF}
\end{equation}
where $B_\textsc{IMF}=\left|\mathbf{B}_\text{IMF}\right|$ and
$\theta_\textsc{IMF}$ is the clock angle that the IMF makes with the
$z$-axis.  The positions $\mathbf{r}_\text{Null}$ of the magnetic
nulls satisfy
\begin{equation}
\mathbf{B}_\text{D}\left(\mathbf{r}_\text{Null}\right)+\mathbf{B}_\text{IMF}=0.
\label{eqn::VS_equal_0}
\end{equation}
For the chosen form of $\mathbf{B}_\text{IMF}$ and using no dipole
tilt, the nulls in spherical coordinates are at
$\mathbf{r}_\text{Null}=\left({r}_\text{Null},\,~\phi_\text{Null}=\pm90^\circ,\,~\pm\lambda_\text{Null}\right)$~\citep{yeh1976,hu2009},
where
\begin{equation}
r_\text{Null}=\left(\frac{M}{2B_\textsc{IMF}}\right)^{1/3}\left[\cos{\theta_\textsc{IMF}}+\sqrt{8+\cos^2{\theta_\textsc{IMF}}}\right]^{1/3}
\label{eqn::Null_Radius}
\end{equation}
and 
\begin{equation}
\lambda_\text{Null}=\tan^{-1}{\left(\frac{3\cos{\theta_\textsc{IMF}}+\sqrt{8+\cos^2{\theta_\textsc{IMF}}}}{4\sin{\theta_\textsc{IMF}}}\right)}.
\label{eqn::Null_Latitude}
\end{equation}
In GSE coordinates, $\mathbf{r}_\text{Null}=(x_\text{Null},\,~y_\text{Null},\,~z_\text{Null})$ 
with
\begin{equation}
  x_\text{Null}=0,
\label{eqn::x_null}
\end{equation}
\begin{equation}
  y_\text{Null}=r_\text{Null}\left(\frac{3+\sin^2\theta_\text{IMF}-\cos\theta_\text{IMF}\sqrt{\cos^2\theta_\text{IMF}+8}}{6}\right)^{1/2},
\label{eqn::y_null}
\end{equation}
and
\begin{equation}
  z_\text{Null}=r_\text{Null}\left(\frac{2+\cos^2\theta_\text{IMF}+\cos\theta_\text{IMF}\sqrt{\cos^2\theta_\text{IMF}+8}}{6}\right)^{1/2}.
\label{eqn::z_null}
\end{equation}
The nulls lie in the dawn-dusk plane $\left(x=0\right)$, as there is
no $B_x$ component to the IMF field. The separator is a semicircular arc 
of radius $r_\text{Null}$
connecting the two nulls~\citep{cowley1973,yeh1976,hu2009}.  To motivate that this is the case, note that for pure southward $B_\text{IMF}$, the separator is a circle in the ecliptic plane.  For other clock angles, the separator rotates out of the ecliptic plane by $\lambda_\text{Null}$ without changing its shape.

\begin{figure}
\centering
\noindent\includegraphics[width=20pc]{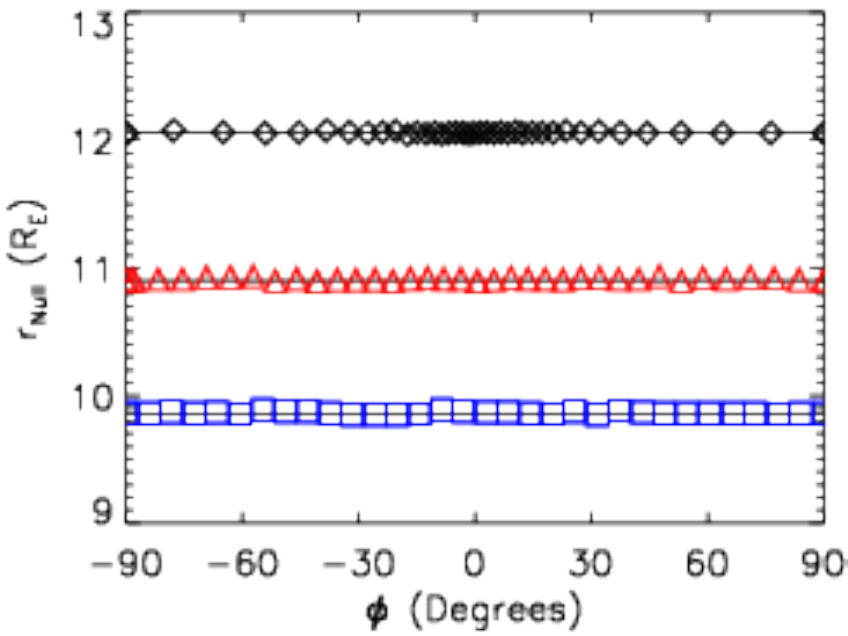}
\caption{Verification of the separator tracing algorithm for
  $\theta_\textsc{IMF}=30^\circ$ (black diamonds),~$90^\circ$ (red
  triangles), and $150^\circ$ (blue squares) using vacuum
  superposition. The exact solutions for the separator are shown as
  solid (black) lines.}
\label{fig::Separator_Characterization}
\end{figure}

We use the technique from Sec.~\ref{subsection::Technique_Description}
to trace separators in vacuum superposition with
$\theta_\textsc{IMF}~=~30^\circ,\,~90^\circ,\,~\text{and
}\,~150^\circ$ for a system with $\mathbf{M}~=~-~5.13~\times~10^4\
~\text{nT}\ ~\text{R}_\text{E}^3\ ~\mathbf{\hat{z}}$ and
$B_\textsc{IMF}=56~\ \text{nT}$. The tracing algorithm uses
hemispheres with radii $R_\text{HS}=1~\ \text{R}_\text{E}.$ The
separator locations are plotted in
Fig.~\ref{fig::Separator_Characterization} as (black) diamonds for
$\theta_\text{IMF}=30^\circ$, (red) triangles for $90^\circ$, and
(blue) squares for $150^\circ$. The exact solutions for the separator
from Eq.~(\ref{eqn::Null_Radius}) are plotted as solid (black)
lines. The measured separator locations agree exceedingly well with
the exact solutions.

To test the accuracy of the algorithm, we repeat the tracing using
hemispheres with radii $R_\text{HS}=5~\ \text{R}_\text{E}$ (not
shown). As expected, the agreement is better with hemispheres of
smaller radii. The scatter of the separator locations from the exact
solution in Eq.~(\ref{eqn::Null_Radius}), measured as the average
absolute difference between the measured separator radius and
$r_\text{Null}$, is $\sim75\%$ lower when smaller hemispheres are
used.

\section{Magnetospheric Simulation Study}
\label{section::Simulation_Study}
\subsection{The Code and its Initialization}
\label{subsection::Code_Initialization}
To find separators on a self-consistently generated magnetosphere,
global simulations using the Block Adaptive Tree Solarwind Roe-Upwind
Scheme (BATS-R-US)~\citep{powell1999,gombosi2000,dezeeuw2000} are
performed at NASA's CCMC. BATS-R-US solves the MHD equations on a
three-dimensional rectangular irregular grid. The simulation domain is
$-255<x<33,-48<y<48,-48<z<48$, where distances are measured in
$\text{R}_\text{E}$ and the coordinate system is GSE.

The simulations are run using BATS-R-US version 8.01 and do not use
the Rice Convection Model (RCM). The simulations are evolved for two
hours (02:00:00) of magnetospheric time. We look at the 02:00:00 mark
of simulation data because the system has achieved a quasi-steady
state; the magnetopause current layer along the $x$-axis is
approximately stationary. The standard high-resolution grid for CCMC
simulations has $1,958,688$ grid cells with a coarse resolution of $8\
~\text{R}_\text{E}$ in the far magnetotail, and a fine resolution of
$0.25\ ~\text{R}_\text{E}$ near the magnetopause. The present study
employs a higher resolution grid of $0.125\ ~\text{R}_\text{E}$ packed
in the region $-6 < x < 10, -10 < y < 10, -5 < z < 5$
$\text{R}_\text{E}$ with $3,736,800$ total grid cells.

The simulations do not employ a dipole tilt and use fixed solar wind
inflow conditions. The solar wind has temperature $T=232,100\
~\text{K}$, IMF strength $20\ ~\text{nT}$, number density $n=20\
~\text{cm}^{-3}$, and a solar wind velocity of ${\bf v}=-400\
~\text{km/s}\ ~\mathbf{\hat{x}}$. We perform distinct simulations with
IMF clock angle $\theta_\text{IMF}=0^\circ$ (parallel), $30^\circ$,
$60^\circ$, $90^\circ$, $120^\circ$, $150^\circ$, $165^\circ$, and
$180^\circ$ (anti-parallel). The IMF does not have a $B_x$
component. Constant Pederson and Hall conductances of 5 mho are
used. The solar radio flux F10.7 index is set at a value of 150.

\subsection{Numerical  vs. Explicit Dissipation}
\label{subsection::Numerical_Dissipation}
\begin{figure*}
\centering
\noindent\includegraphics[width=39pc]{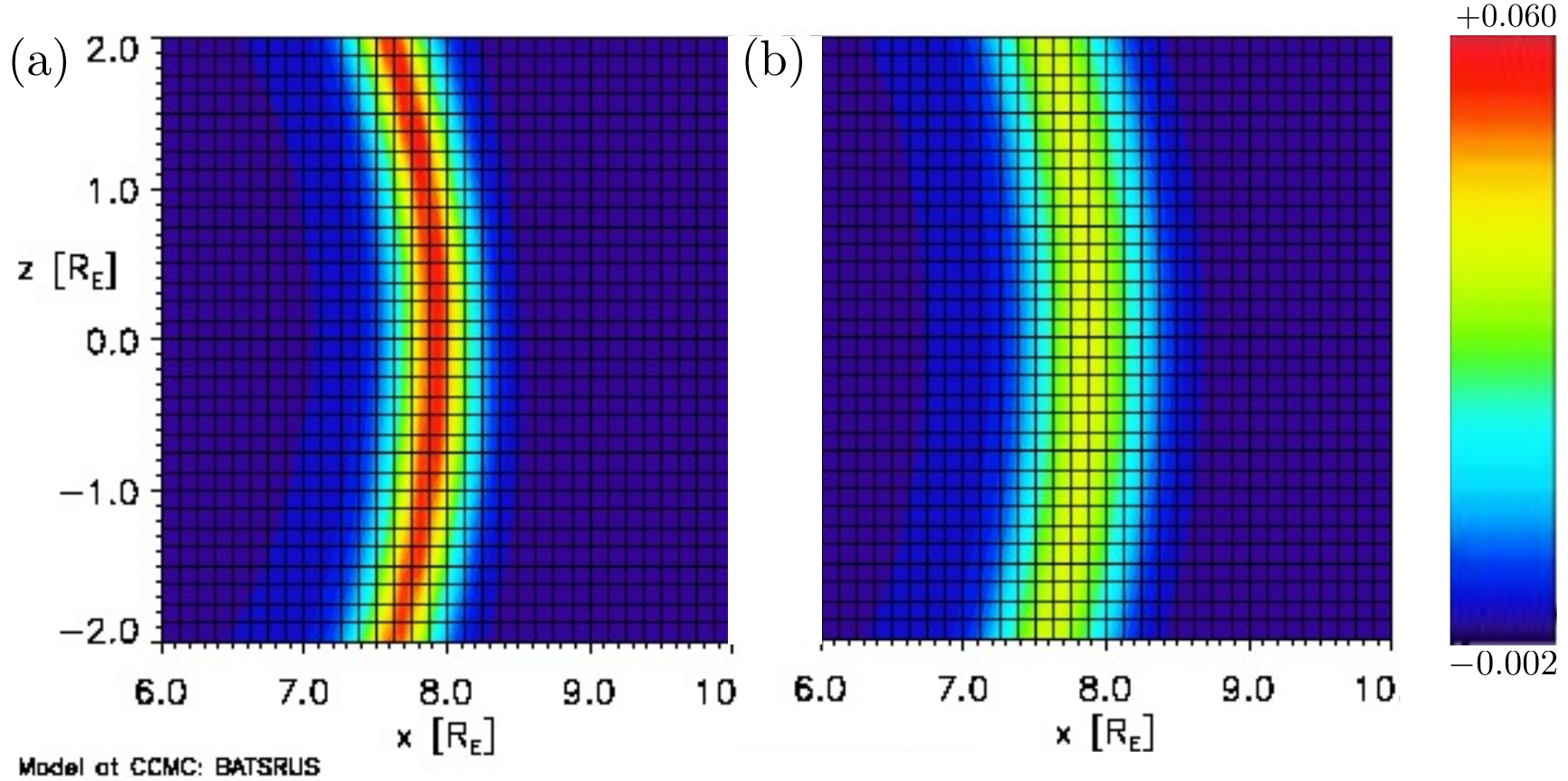}
\caption{Magnetopause current density
  $J_y\,~\left(\mu\text{A}/\text{m}^2\right)$ near the subsolar point
  in high resolution simulations (a) without an explicit resistivity
  and (b) with an explicit resistivity $\eta/\mu_0=6.0~\times10^{10}~\
  \text{m}^2/\text{s}$. The current layer broadens in (b),
  demonstrating the explicit resistivity dominates numerical
  effects. Solid (black) lines indicate the simulation grid
  $\left(0.125\ ~\text{R}_\text{E}\right)$.}
\label{fig::Resistivity_Comparison}
\end{figure*}

Many global MHD simulations use numerical grid-scale dissipation
instead of explicit dissipation because the latter is unnecessary for
large scale physics and can even be detrimental~\citep{raeder1999}.
However, explicit dissipation is essential for studies involving
magnetic reconnection and magnetic topology.  For example, global
simulations revealed plasma flows through the reconnection
X-line~\citep{siscoe2002,dorelli2004}; some researchers believed this
to be spurious due to high dissipation, but later studies showed this
flow is a fundamental aspect of asymmetric
reconnection~\citep{cassak2007,cassak2009c}.

For the present simulations, we employ a uniform explicit resistivity
$\eta$.  It is known that the magnetosphere is not collisional, but
including an explicit resistivity allows for reproducible results that
are independent of the numerics. We include an explicit resistivity
$\eta/\mu_0=6.0~\times~10^{10}\ ~\text{m}^2/\text{s}$ in our
simulations.  To ensure this resistivity controls the dissipation
instead of the numerics, we show $J_{y}$ in the $y = 0$ plane at the
dayside magnetopause in Fig.~\ref{fig::Resistivity_Comparison} for (a)
no explicit resistivity and (b) with the explicit resistivity.  The
current layer broadens from six cells to eight cells across,
suggesting the explicit resistivity is controlling the dissipation, as
desired.  This explicit resistivity is similar to the value obtained
in a recent study that determined the size of the resistivity
necessary for it to control the dissipation (G. Toth, private
communication).

To facilitate comparisons with previous simulations, we estimate a
Lundquist number $S=\mu_0 c_{A} L/\eta$ for the explicit resistivity
simulation with $\theta_\text{IMF}=180^\circ$. We base the length
scale $L$ on the half-length of the reconnecting current sheet in the
outflow direction (as opposed to a global length scale), which is
$5.35\ ~\text{R}_\text{E}$. For the Alfv\'en speed $c_{A}$, we use a
hybrid magnetosheath/magnetospheric value $c_{Ah}$ based on the
asymmetric reconnection theory of~\citet{cassak2007} of the form
\begin{equation}
  c_{Ah}^2\sim\frac{B_1B_2\left(B_1+B_2\right)}{\mu_0\left(\rho_1B_2+\rho_1B_2\right)},
  \label{eqn::Asymmetric_Alfven}
\end{equation}
where $B$ and $\rho$ are the magnetic fields and plasma densities
measured upstream of the current layer, and subscripts ``1'' and ``2''
indicate quantities measured in the magnetosphere and
magnetosheath. The magnetic fields and densities, measured immediately
upstream of the reconnecting current sheet in the Earthward and
Sunward directions, are $B_1=116$ nT, $B_2=90$ nT, $n_1=10\
~\text{cm}^{-3}$ and $n_2=57\ ~\text{cm}^{-3}$, giving
$c_{Ah}\simeq380\ ~\text{km/s}$. The resulting Lundquist number based
on these quantities and our chosen explicit resistivity is
$S\simeq210$. A benefit of choosing the explicit resistivity this
large is that the rate of production of plasmoids ({\it i.e.,} flux
transfer events - FTEs) is decreased.  While FTEs do occur at the
magnetopause, they would needlessly complicate the present fundamental
physics study on field line topology.

\section{Results}
\label{section::Results}
\begin{table}
  \caption{The $(x,\,~y,\,~z)$ coordinates (in GSE) of determined 
    magnetic nulls in global magnetosphere simulations and in vacuum 
    superposition with $\mathbf{M}~=~-~5.13~\times~10^4\ ~\text{nT}\ ~\text{R}_\text{E}^3\ ~\mathbf{\hat{z}}$ and $B_\textsc{IMF}=56~\ \text{nT}$.}
\centering
\begin{tabular}{ c c c}
  \hline
  Clock Angle & MHD Nulls $\left(\text{R}_\text{E}\right)$ & Vacuum Nulls $\left(\text{R}_\text{E}\right)$\\\hline
  $0^\circ$ & $( 0.08,\,0.00,\,\pm10.28)$ & $(0.00,\,0.00,\,\pm12.24)$\\\hline
  $30^\circ$ & $(-0.10,\,\pm2.99,\,\pm9.99)$ & $(0.00,\,\pm4.08,\,\pm11.34)$\\\hline
  $60^\circ$ & $(-0.44,\,\pm5.19,\,\pm9.35)$ & $(0.00,\,\pm7.18,\,\pm9.06)$\\\hline
  $90^\circ$ & $(-0.41,\,\pm7.70,\,\pm7.96)$ & $(0.00,\,\pm8.91,\,\pm6.29)$\\\hline
  $120^\circ$ & $(-0.92,\,\pm9.55,\,\pm5.83)$ & $(0.00,\,\pm9.56,\,\pm3.78)$\\\hline
  $150^\circ$ & $(-2.16,\,\pm11.03,\,\pm3.46)$ & $(0.00,\,\pm9.70,\,\pm1.75)$\\\hline
  $165^\circ$ & $(0.93,\,\pm10.11,\,\pm1.89)$ & $(0.00,\,\pm9.71,\,\pm0.85)$\\\hline
\end{tabular}
\label{table::Nulls}
\end{table}

Here, we describe the results of finding nulls and tracing separators
in the global, resistive MHD simulations described in
Sec.~\ref{section::Simulation_Study}. To develop perspective on the
results, we compare the results to nulls and separators in vacuum
superposition given by Eqs.~(\ref{eqn::Null_Radius})
and~(\ref{eqn::Null_Latitude}). To make a careful comparison, we do
not use the nominal values of $\mathbf{M}~=~-~3.11~\times~10^4\
~\text{nT}\ ~\text{R}_\text{E}^3\ ~\mathbf{\hat{z}}$ and
$B_\textsc{IMF}=20~\ \text{nT}$ because the terrestrial magnetic field
is enhanced due to compression by the solar wind and the IMF increases
at the bow shock. We find more appropriate values from the MHD
simulations. The magnetic field strengths are measured upstream of the
current sheet at the subsolar point. To do so, the locations where the
magnetopause current drops to $1/e$ of its maximum on the sun-Earth
line is found for both sides of the sheet. On the magnetospheric side,
the magnetic field averages a 65\% increase over Earth's nominal
dipole field in our simulations for all clock angles, so we employ
$\mathbf{M}=-5.13\times10^4~\ \text{nT}\ ~\text{R}_\text{E}^3~\
\hat{\mathbf{z}}$. On the magnetosheath side, $B_\text{IMF}\simeq56\
~\text{nT}$ for all clock angles. These are the values we employ for
the vacuum superposition fields.

\subsection{The Magnetic Nulls}
\label{subsection::Magnetic_Nulls}

\begin{figure*}
\centering
\noindent\includegraphics[width=39pc]{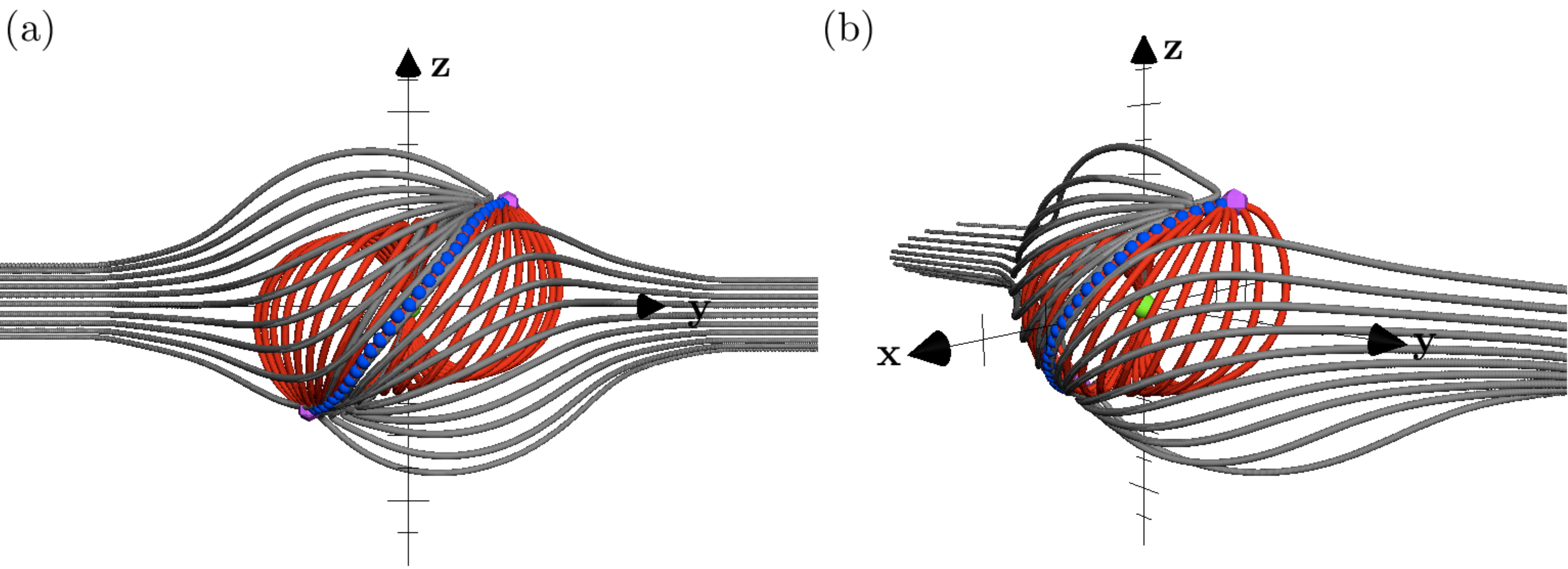}
\caption{Results of the present algorithm in the MHD simulation with $\theta_\text{IMF}=90^\circ$ looking (a) Earthward and (b) at an oblique angle for perspective.  Magnetic nulls are enclosed by (purple) spheres and Earth is the (green) sphere at the origin (to scale).  The last closed magnetic field lines in the ecliptic plane are displayed in red and adjacent half-closed topologies are displayed in gray.}
\label{fig::90_Nulls}
\end{figure*}

We find the magnetic nulls by employing the~\citet{haynes2007a} method
and their GSE locations are listed in Table~\ref{table::Nulls} for the
MHD simulations and for the vacuum superposition fields. The latter
are in close agreement with
Eqs.~(\ref{eqn::x_null})$-$(\ref{eqn::z_null}).  The measured magnetic
field strength at each of the locations identified as nulls is $0.1~\ \text{nT}$ or
lower. Nulls for $\theta_\textsc{IMF}=180^\circ$ are not reported as
there are an infinite number of them in the ecliptic plane.  As an example, the magnetic nulls for the $\theta_\text{IMF}=90^\circ$ MHD simulation are plotted as purple spheres in Fig.~\ref{fig::90_Nulls}, showing an (a) Earthward and (b) oblique view for perspective.  The Earth is depicted as the green sphere (to scale).

The location of the magnetic nulls exhibit a few interesting trends as
a function of IMF clock angle. The nulls in vacuum superposition have
a range of $12.3\ge r_\text{Null}\ge 9.7\ ~\text{R}_\text{E}$ for
$0^\circ\le\theta_\text{IMF}\le 180^\circ$, whereas the nulls found in
MHD have a nearly constant $r_\text{Null}\simeq10.5\
~\text{R}_\text{E}$. The trend in $r_{Null}$ differs because the
magnetopause is located where the magnetospheric magnetic pressure
balances the solar wind ram pressure in MHD. Our MHD simulations all
have the same solar wind conditions, which explains why
$r_\text{Null}$ remains constant in MHD. Vacuum superposition is only
a magnetic field model and does not capture this solar wind
physics. The magnetopause shrinks as $\theta_\text{IMF}$ increases
because the $B_z$ component of $B_\text{IMF}$ becomes increasingly
negative, enabling the IMF to penetrate further into the
magnetosphere.

\begin{figure}
\centering
\noindent\includegraphics[width=20pc]{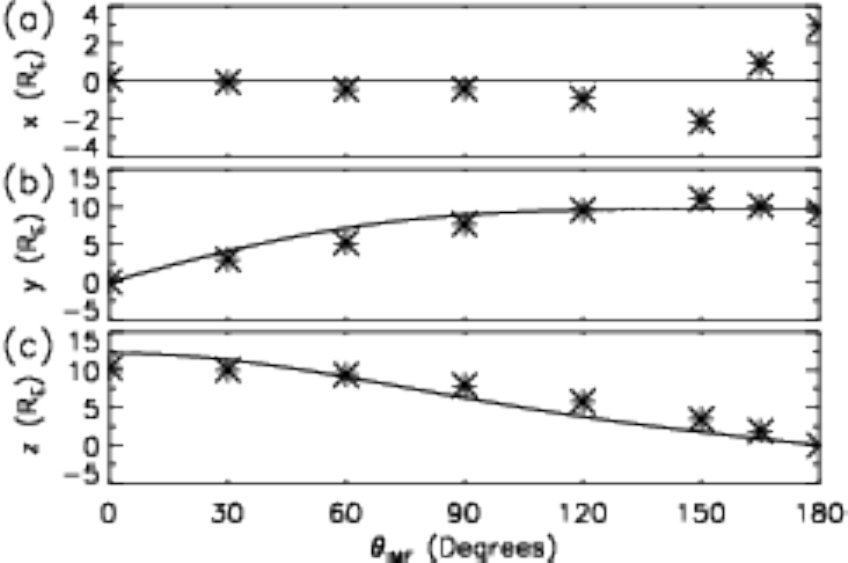}
\caption{Plots of magnetic null (a) $x$-coordinate, (b)
  $y$-coordinate, and (c) $z$-coordinate as a function of IMF clock
  angle $\theta_\text{IMF}$. Solid lines display vacuum superposition
  prediction from Eqs.~(\ref{eqn::x_null})$-$(\ref{eqn::z_null}) and
  asterisks are the coordinates of nulls in the MHD simulations.}
\label{fig::Null_Plots}
\end{figure}

Figure~\ref{fig::Null_Plots} displays the measured MHD coordinates of
the nulls as asterisks and the solid lines as the predicted values for
vacuum superposition from
Eqs.~(\ref{eqn::x_null})$-$(\ref{eqn::z_null}) as a function of
$\theta_\text{IMF}$. The $y$- and $z$-coordinates of the nulls follow
qualitatively similar trends for both vacuum superposition and MHD and
are within $2\ ~\text{R}_\text{E}$ of each other for all clock
angles. In Fig.~\ref{fig::Null_Plots}(b), the $y$-coordinate increases
from zero as $\theta_\text{IMF}$ increases, and the nulls move out of
the noon-midnight plane. The $z$-coordinate decreases to zero for
increasing clock angle, as seen in Fig.~\ref{fig::Null_Plots}(c).
This is because the nulls are located at the magnetic cusps for
northward IMF and are in the ecliptic plane for southward IMF.

The $x$-coordinates of the nulls displayed in
Fig.~\ref{fig::Null_Plots}(a) do not follow the same trend as in
vacuum superposition. The nulls in vacuum superposition are in the
dawn-dusk plane $\left(x=0\right)$ for all $\theta_\text{IMF}$ as
there is no $B_x$ component of the IMF [see
Eq.~(\ref{eqn::x_null})]. In the MHD simulations, the nulls are near
$x=0$ for small clock angles, but migrate towards the nightside as
$\theta_\text{IMF}$ increases towards $150^\circ$. Interestingly, this
trend is broken for $\theta_\text{IMF}=165^\circ$ which has a null
with a $+x$-coordinate.

One might suggest the migration of the nulls to the nightside results from the draping of the IMF over the magnetosphere.  Draping causes the IMF to be oriented sunward in the southern hemisphere and tailward in the northern hemisphere for northward IMF, with the opposite being true for southward IMF.  However, this effect would make the nulls migrate opposite to the observed direction, so draping cannot explain the migration of the nulls' $x$-coordinate.  We conclude that there is no simple explanation of the trend in the $x$-coordinate of the nulls, but this is not surprising since null locations are dependent on the shape of the magnetopause, which has a multi-parameter dependence on upstream solar wind conditions~\citep{lu2011,liu2012}.  

\subsection{The Magnetic Separators}
\label{subsection::The_Separators}
The separator tracing method described in
Sec.~\ref{section::Finding_Separators} is used to trace the dayside
separators for the MHD simulations. We start from the magnetic nulls
described in the previous section and use hemispheres with radii of
$R_\text{HS}=1~\ \text{R}_\text{E}$ to trace the separators.

Care must be taken in tracing the separator for
$\theta_\text{IMF}=180^\circ$ due to the infinite number of nulls in
the ecliptic plane. We start by centering a sphere at the subsolar
point $\mathbf{r}_\text{Null}=(7.87,0.00,0.00)\
~\text{R}_\text{E}$. We center the hemisphere at the subsolar point,
and the hemisphere is discretized into the same $N_\phi\times
N_\lambda$ grid as described in
Section~\ref{subsection::Technique_Description}. The hemisphere's
coordinates span longitude $0^\circ\le\phi\le180^\circ$ and latitude
$-90^\circ\le\lambda\le90^\circ$. The chosen longitude range only
traces the portion of the separator duskward of the subsolar
point. The algorithm iteratively marches in the ecliptic plane until
it no longer detects a merging location, ending at
$\mathbf{r}=(2.93,9.33,0.00)\ ~\text{R}_\text{E}$. The dawnward
portion of the separator is traced likewise by forcing the hemisphere
to have a longitude range of $-180^\circ\le\phi\le0^\circ$, ending at
$\mathbf{r}=(2.93,-9.33,0.00)\ ~\text{R}_\text{E}$. The resulting
separator is stitched together with the subsolar point as the center
of each portion.  

An example of a traced separator is shown in Fig.~\ref{fig::90_Nulls}, with the blue spheres denoting the intersection of the separator with the hemispheres form the iterative technique described in Sec.~\ref{subsection::Technique_Description}.  For perspective, the last closed field lines in the ecliptic plane are shown in red, and the adjacent half-closed field lines are shown in gray.\newline

\subsubsection{Comparison with the Last Closed Field Line}
\label{subsubsection::Last_Closed_Comparison}
\begin{figure}
\centering
\noindent\includegraphics[width=20pc]{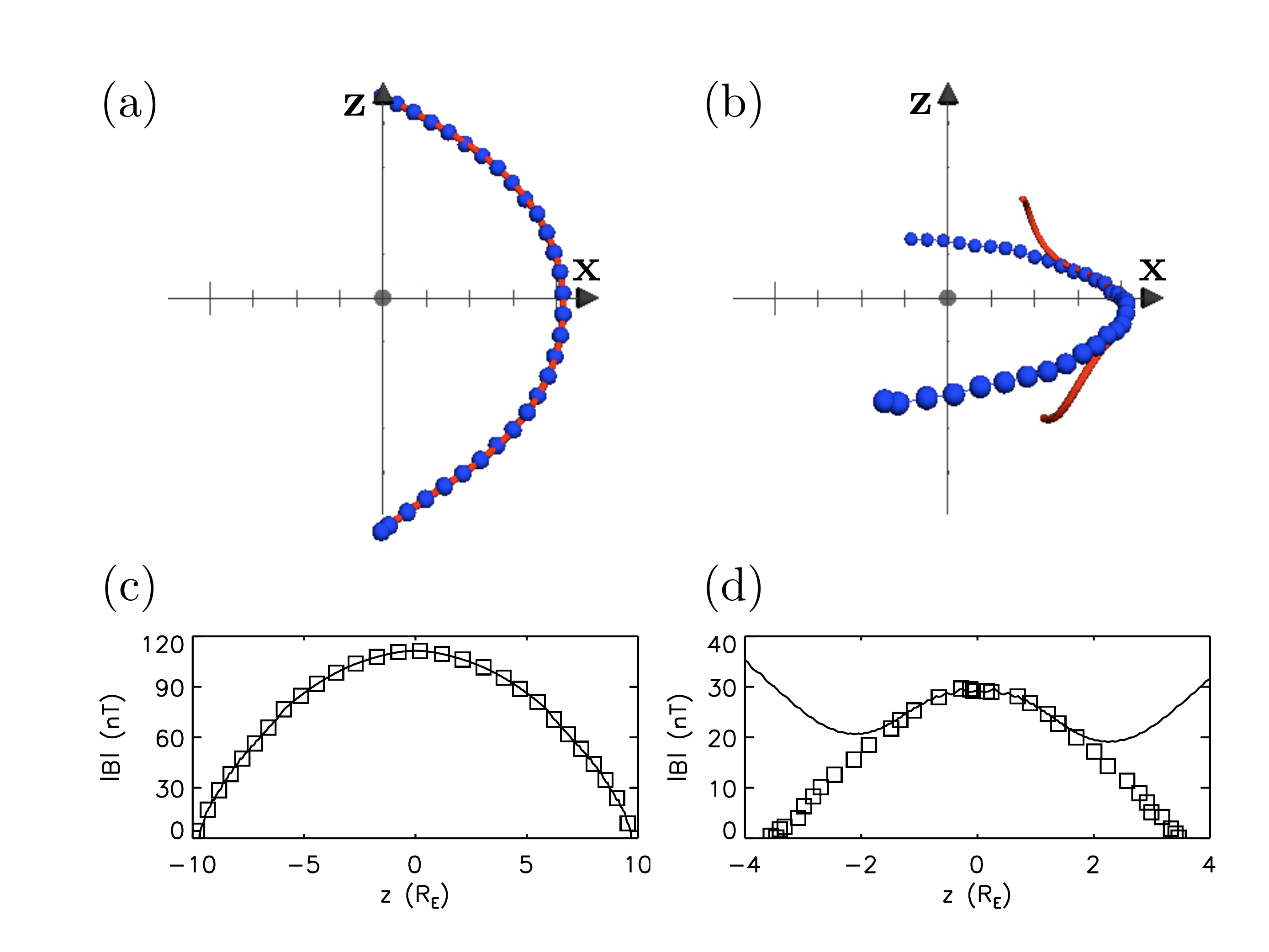}
\caption{Comparison between traced separators and the last closed
  field line in global MHD simulations. The last closed field line is
  a solid (red) line and the determined separator locations are (blue)
  spheres for $\theta_\text{IMF}$ of (a) $30^\circ$ and (b)
  $150^\circ$. The magnetic field strength $\left|\mathbf{B}\right|$
  as a function of $z$-coordinate along the separator is a solid line
  for the last closed field line and as squares for the traced
  separator for $\theta_\text{IMF}$ (c) $30^\circ$ and (d)
  $150^\circ$.}
\label{fig::Last_Closed_Comparison}
\end{figure}

The last closed field line on the sun-Earth line has been used to
approximate the magnetic separator since it closely approaches both
magnetic nulls (northward IMF:~\citep{dorelli2007}; southward and
northward IMF:~\citet{hu2009}). We compare the traced separators with
the last closed field lines on the $x$-axis for two different clock
angles. Figures~\ref{fig::Last_Closed_Comparison}(a) and (b) show the
last closed field line as a solid (red) line and the individual
locations determined by the method described in
Sec.~\ref{section::Finding_Separators} as (blue) spheres for
$\theta_\text{IMF}=30^\circ$ and $150^\circ$, respectively. The traced
separator and last closed field line are nearly identical in (a),
where the IMF has a northward $B_z$. In contrast, the two have a large
deviation in (b), where the IMF has a southward $B_z$. Panels (c) and
(d) display $\left|\mathbf{B}\right|$ as a function of separator
$z$-coordinate along the separator, with the last closed field line
shown as a solid (black) line and the locations of the traced
separator plotted as squares for the same two cases. The last closed
field line and the traced separator are coincident in (c) and,
importantly, both connect with the magnetic nulls. In (d), the traced
separator closely agrees with the last closed field line near the
subsolar point, but only the traced separator connects with the
magnetic nulls, while the last closed field line diverges strongly.
More generally, we find that both methods agree near the subsolar
point for all $\theta_\text{IMF}$, but as $\theta_\text{IMF}$
increases from $90^\circ$, the last closed field line increasingly
deviates from the traced separators. Therefore, the last closed field
line does not accurately map the entire separator for southward $B_z$
in our simulations. While the last closed field line is accurate for
northward $B_z$, the method of
Sec.~\ref{subsection::Technique_Description} works for any clock
angle. \newline

\subsubsection{Clock angle dependence of MHD Separators}
\label{subsubsection::Separator_Clock_Angle_Dependence}
\begin{figure}
\centering
\noindent\includegraphics[width=20pc]{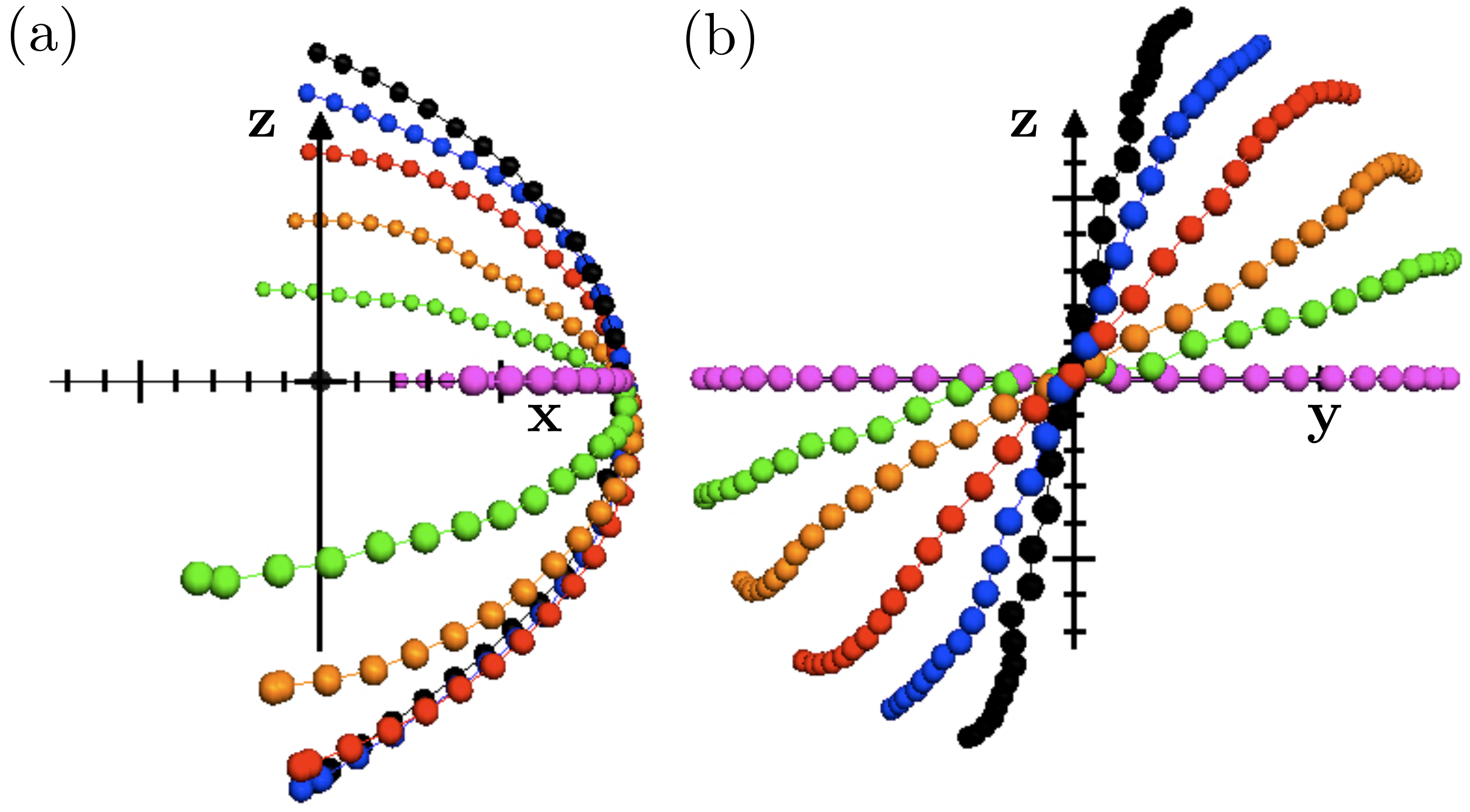}
\caption{Plot of separators in global MHD simulations for
  $\theta_\textsc{IMF}=30^\circ$ (black), $60^\circ$ (blue),
  $90^\circ$ (red), $120^\circ$ (orange), $150^\circ$ (green),
  $180^\circ$ (purple) looking (a) duskward and (b) earthward.}
\label{fig::Separators}
\end{figure}

We now turn to comparing separators for different clock angles. The
separators traced for clock angles $30^\circ$ through $180^\circ$ are
displayed in Fig.~\ref{fig::Separators}. Panel (a) displays the
separators looking duskward along the $y$-axis and (b) displays the
separators looking Earthward along the $x$-axis. Each separator is
roughly coplanar, and is tilted around the $x$-axis by an amount
dependent on the clock angle.

To quantify the structural properties of the separator, we define the
separator tilt angle $\kappa$ at the subsolar point measured with
respect to the $z$-axis. The tilt angles of the separators are
measured using the $B_y$ and $B_z$ components of the last closed field
line at the subsolar point:
\begin{equation}
\kappa_\text{MHD}=\tan^{-1}\left(\frac{B_y}{B_z}\right).
\label{eqn::Separator_Tilt}
\end{equation}
To investigate the separator shape as a function of clock angle, we
rotate clockwise around the $+x$-axis by $\kappa_\text{MHD}$ and
display the separator's projection in this rotated plane. The
separator's projected coordinates in this plane are given by
\begin{equation}
  \left(\begin{array}{c}x'\\y'\\z' \end{array}\right)=
  \left(\begin{array}{c c c} 
      1 & 0 & 0 \\ 
      0 & \cos{\kappa_\text{MHD}} & -\sin{\kappa_\text{MHD}}\\
      0 & \sin{\kappa_\text{MHD}} & \cos{\kappa_\text{MHD}}\\
    \end{array}\right)\left(\begin{array}{c} x \\ y \\ z \end{array}\right),
\label{eqn::Tilted_Plane}
\end{equation}
where $x'$ points sunward, $y'$ is the out-of-plane direction, $z'$ is
the plane of the separator, and $\left(x,\,~y,\,~z\right)$ is the
vector for a given location on the separator in GSE coordinates.

\begin{figure}
\centering
\noindent\includegraphics[width=20pc]{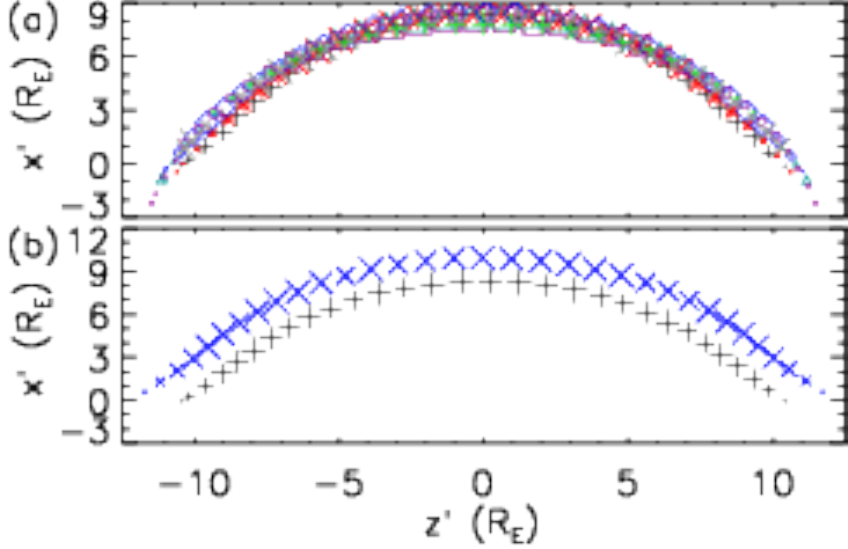}
\caption{MHD separators rotated around $\mathbf{\hat{x}}$ into a
  common plane. (a) Separators for the different clock angles are:
  $30^\circ$ (black plus), $60^\circ$ (red asterisk), $90^\circ$ (blue
  diamond), $120^\circ$ (green triangle), $150^\circ$ (purple
  squares), $165^\circ$ (grey X), and $180^\circ$ (green plus). (b)
  Plot of MHD separators with $\theta_\text{IMF}=30^\circ$ with solar
  wind number density $n=20~\ \text{cm}^{-3}$ (black plus) and $n=5~\
  \text{cm}^{-3}$ (blue X). Symbol size is inversely related to the
  deviation from the plane.}
\label{fig::Separator_Plane}
\end{figure}

Figure~\ref{fig::Separator_Plane}(a) shows the separator's projection
in the rotated plane for different IMF clock angles:
$\theta_\text{IMF}=30^\circ$ as (black) pluses
$\left(\kappa_\text{MHD}\simeq12.6^\circ\right)$, $60^\circ$ as (red)
asterisks $\left(\kappa_\text{MHD}\simeq21.4^\circ\right)$, $90^\circ$
as (blue) diamonds $\left(\kappa_\text{MHD}\simeq40.2^\circ\right)$,
$120^\circ$ as (green) triangles
$\left(\kappa_\text{MHD}\simeq62.8^\circ\right)$, $150^\circ$ as
(purple) squares $\left(\kappa_\text{MHD}\simeq79.4^\circ\right)$,
$165^\circ$ as (gray) X's
$\left(\kappa_\text{MHD}\simeq79.0^\circ\right)$, and $180^\circ$ as
(green) pluses $\left(\kappa_\text{MHD}=90^\circ\right)$. The symbol
size denotes the location's deviation from the plane, with smaller
symbols indicating a larger deviation from the plane.

Figure~\ref{fig::Separator_Plane}(a) simultaneously quantifies three
structural features of the magnetic separators. It is clearly seen
that the separators maintain a similar shape regardless of
$\theta_\text{IMF}$. Also, the separators rotate around the
magnetopause for increasing $\theta_\text{IMF}$, turning clockwise
around the $x$-axis.  Finally, the symbols indicate that a large
portion of the separator is approximately coplanar in the plane
defined by $\kappa_\text{MHD}$, particularly at the nose of the
magnetosphere where the deviation from the plane is $\le0.2~\
\text{R}_\text{E}$. The deviation is larger near the nulls
($\simeq1.5~\ \text{R}_\text{E}$), which can clearly be seen in
Fig.~\ref{fig::Separators}(b), where the ends of the separators flare
towards the dawn and dusk flanks. This implies that it is not accurate
to model separators as lying in the plane of the nulls.\newline

\subsubsection{Comparison with Vacuum Superposition}
\label{subsubsection::Separator_Comparison_with_Vacuum_Superposition}
To gain perspective on the observed trends in separators in MHD
simulations with varying IMF clock angle, we compare them to vacuum
superposition separators, although a perfect correlation is not
expected. As discussed earlier, MHD separators are mostly coplanar;
the vacuum superposition separators are exactly coplanar. Also, the
shape of the separator is different between the models. In vacuum
superposition, the separator is a circular arc with radius
$r_\text{Null}$ given by Eq.~(\ref{eqn::Null_Radius}). The MHD
separators exhibit the well known bullet shape of the magnetopause, as
seen in Fig.~\ref{fig::Separator_Plane}.

\begin{figure}
\centering
\noindent\includegraphics[width=20pc]{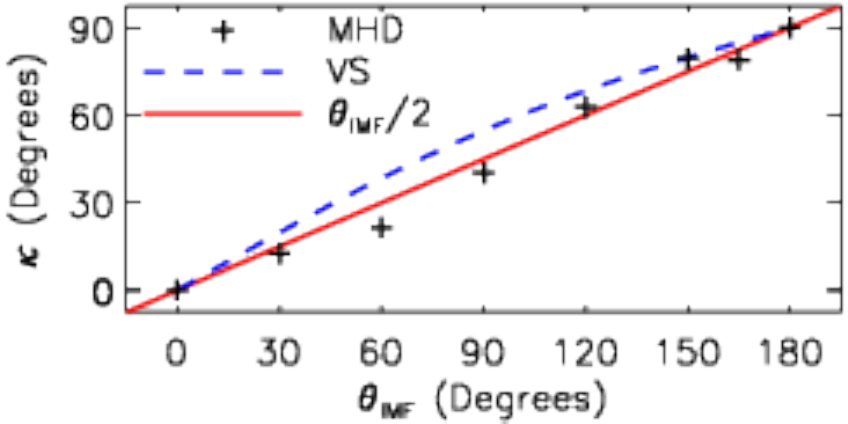}
\caption{Separator tilt angle $\kappa$ at the subsolar point as a
  function of IMF clock angle $\theta_\text{IMF}$. The solid (red)
  line is the bisection angle $\theta_\text{IMF}/2$, the dashed (blue)
  line is for vacuum superposition separators and the pluses are for
  MHD separators.}
\label{figure::Theta_v_Clock_Angle}
\end{figure}

The separator tilt angle dependence on IMF clock angle is displayed in
Fig.~\ref{figure::Theta_v_Clock_Angle}. The black pluses display
$\kappa_\text{MHD}$ as calculated by Eq.~(\ref{eqn::Separator_Tilt}).
For the vacuum superposition separators,~\citet{yeh1976}
and~\citet{hu2009} showed the separator tilt angle satisfies
$\kappa_\text{VS}=90^\circ-\lambda_\text{Null}$, with
$\lambda_\text{Null}$ given by Eq.~(\ref{eqn::Null_Latitude}),
displayed as the dashed (blue) line. Lastly, the solid (red) line
shows $\kappa=\theta_\text{IMF}/2$, the angle bisecting the IMF and
terrestrial magnetic field, a commonly used estimate, for reference.

The separator tilt angle $\kappa$ increases from $0^\circ$ to
$90^\circ$ for vacuum superposition, MHD, and angle of bisection. The
tilt angles for the three $\kappa$ values are relatively close to each
other, within about $20^\circ$. However, quantitative predictions and
trends with clock angle reveal important differences between the
models. In vacuum superposition, the tilt angle is consistently larger
than the angle of bisection, implying separators that are tilted
towards the ecliptic plane. MHD separators are tilted towards the
noon-midnight meridional plane for $\theta_\text{IMF}\le90^\circ$, but
are tilted towards the ecliptic for $\theta_\text{IMF}=120^\circ$ and
$150^\circ$. Therefore, the three models follow similar trends at
small and large IMF clock angles, but the MHD separator tilt angle
displays significant differences from the models for intermediate
clock angles.  \newline

\subsubsection{Density Dependence of Separators}
\label{subsubsection::Separator_Density_Dependence}
As a preliminary test of the parametric dependence of MHD separator
characteristics, we perform a simulation similar to our
$\theta_\text{IMF}=30^\circ$ simulation, only changing solar wind
number density to $n=5\ ~\text{cm}^{-3}$ (from $n=20\
~\text{cm}^{-3}$). We expect the magnetosphere to expand with this
decrease in number density. The location of the magnetopause
$R_\text{MP}$ on the $x$-axis occurs approximately where the solar
wind dynamic pressure balances the magnetosphere's magnetic
pressure. As the magnetospheric magnetic field is dipolar with
$B_\text{D}\propto1/{r^3}$, the magnetopause location $R_\text{MP}$
depends on density as $R_\text{MP}\propto n^{-1/6}.$ For these
simulations, this implies the magnetopause should approximately expand
by a factor of $\left(5/20\right)^{-1/6}\simeq1.26$. The measured
values of $R_\text{MP}$ from the simulations are $9.94\
~\text{R}_\text{E}$ for $n=5\ ~\text{cm}^{-3}$ and $8.32\
~\text{R}_\text{E}$ for $n=20\ ~\text{cm}^{-3}$, giving a ratio of
$\simeq1.20$, in good agreement with expectations.

The magnetic nulls of the $n=5\ ~\text{cm}^{-3}$ simulation are
located at
$\mathbf{r}_\text{Null}=\left(0.56,\,\pm3.16,\,\pm11.30\right)~\
\text{R}_\text{E}.$ The magnetic nulls of the lower density run are
sunward and radially outward from their high density
counterparts. This behavior is due to the expansion of the
magnetosphere with a lower solar wind density. Thus, the magnetic null
locations are sensitive to the solar wind density.

The expansion of the magnetosphere also affects separator
location. Figure~\ref{fig::Separator_Plane}(b) displays the separators
for the two simulations rotated into the principle plane of the
separator. The original high density run is displayed as (black)
pluses and the low density run as (blue) X's.  As expected, the $n=5\
~\text{cm}^{-3}$ separator expands outward in the $x'$-$z'$ plane. We
measure the separator tilt angle using
Eq.~(\ref{eqn::Separator_Tilt}); the $n=5\ ~\text{cm}^{-3}$ run has
$\kappa_\text{MHD}\simeq12.9^\circ$ and $n=20\ ~\text{cm}^{-3}$ has
$\kappa_\text{MHD}\simeq12.6^\circ.$ Interestingly, despite the
expansion of the magnetosphere changing the separator's location, the
separator's tilt angle is not strongly dependent on solar wind number
density (for the chosen set of simulation parameters). A more complete
parametric study to obtain trends in separator morphology is
necessary.\newline

\subsubsection{Dependence on Grid Resolution}
\label{subsubsection::Separator_Resolution_Dependence}
The null locations given in Table~\ref{table::Nulls} for small and
large IMF clock angle exist near or outside the specified high
resolution grid region given in
Section~\ref{section::Simulation_Study}. We test the dependence of the
nulls and separators on grid resolution by performing additional
simulations with $\theta_\text{IMF}=60^\circ$, $150^\circ$, and
$165^\circ$. The same simulation parameters described in
Section~\ref{section::Simulation_Study} are used, except the $0.125\
~\text{R}_\text{E}$ resolution region with
$\theta_\text{IMF}=60^\circ$ spans $-6<x<10,\,-10<y<10,\,-12<z<12\
~\text{R}_\text{E}$ and $-6<x<10,\,-15<y<15,\,-10<z<10\
~\text{R}_\text{E}$ for the two southward IMF simulations.

The nulls in the higher resolution simulations are located at
$\mathbf{r}_\text{Null}~=~(-0.19,\,~\pm5.23,\,~\pm9.63)\
~\text{R}_\text{E}$ for $60^\circ$,
$\mathbf{r}_\text{Null}~=~(-1.58,\,~\pm11.12,\,~\pm3.20)\
~\text{R}_\text{E}$ for $150^\circ$, and
$\mathbf{r}_\text{Null}~=~(0.30,\,~\pm10.62,\,~\pm2.12)\
~\text{R}_\text{E}$ for $165^\circ$. The nulls in the higher
resolution simulations are $\simeq3$ high resolution grid cells away
from their lower resolution counterparts for $60^\circ$, $\simeq5$
high resolution grid cells for $150^\circ$, and $\simeq7$ high
resolution grid cells for $165^\circ$.  The location of the subsolar
point (as measured by the last closed field line on the $x$-axis) is
$x=8.62\ ~\text{R}_\text{E}$ in the $60^\circ$ higher resolution
simulation, compared to $x=8.44\ ~\text{R}_\text{E}$ for the lower
resolution simulation, a difference of about 1.5 grid cells; the last
closed field lines for $150^\circ$ and $165^\circ$ are within a grid
cell of their lower resolution counterparts. We trace separators in
all higher resolution simulations and find that the separators in the
higher resolution simulations do not deviate significantly from the
lower resolution separators (not shown). This motivates that the
resolution is sufficient to obtain accurate null locations and
separators.

\section{Conclusions}
\label{section::Conclusions}
In summary, we present a simple, efficient, and accurate method of
tracing magnetic separators in global magnetospheric simulations with
arbitrary IMF clock angle. The method is to start at a magnetic null
and iteratively trace the dayside separator by calculating the
magnetic topology on the surface of spherical shells to locate regions
of topological merging. We verify the method using a simple magnetic
field model with exact solutions for the separators. The technique
improves on previous ones by being efficient, good to arbitrary
accuracy, and works for any IMF clock angle.

We then trace separators in several distinct resistive global MHD
simulations with $\theta_\text{IMF}$ ranging from $0^\circ$ to
$180^\circ$. The resulting magnetic nulls and separators in MHD are
compared to those in vacuum superposition. We find that the $y$- and
$z$-coordinates of the magnetic nulls display similar qualitative
trends in both models, but the migration of the null's $x$-coordinate
in MHD is not captured by the vacuum superposition fields. We find
that the method described here can trace MHD separators for arbitrary
clock angle, whereas the last closed field line on the sun-Earth line
only works for northward IMF in our simulations. MHD separators
maintain a similar shape regardless of IMF clock angle and a large
portion of the separators are approximately coplanar, however this
plane does not contain the nulls. We find that both models have
separators that change orientation without appreciably changing shape
much. However, trends within models differ significantly. A
preliminary test of the separator's dependence on solar wind number
density $n$ reveals that the null locations and separator location do
depend on number density, but separator orientation does not strongly
depend on number density for our chosen solar wind parameters.

The present study has focused on the dayside magnetopause, but it
could be useful in other contexts. \citet{xiao2006,xiao2007} observed
magnetic nulls and separators at Earth's nightside and it is plausible
that the method described here could locate separators at the
nightside. The method described in the present study could also be
used in studies of the solar corona and other planetary
magnetospheres.

The results here could assist satellites locate magnetic reconnection
at the dayside magnetopause. This would be of particular interest to
NASA's upcoming Magnetospheric Multiscale (MMS) mission and other
existing missions studying magnetic reconnection at the dayside
magnetopause.

This work employed several simplifying assumptions. The present study included an explicit resistivity to limit the effect of numerical dissipation.  However, this is inappropriate for realistic modeling of Earth's magnetosphere.  It would be interesting and important in future work to find separators in simulations with more realistic collisionless dissipation, such as global Hall-MHD or hybrid simulations, and compare them to the separators in resistive-MHD simulations.

We constructed our simulations to not have flux transfer events
(FTEs)~\citep{russell1978}.  \citet{dorelli2008,dorelli2009} showed that the separator splits into multiple branches in the presence of FTEs at the dayside magnetopause.  In order to trace separators in simulations with FTEs, the presently described method would need to be modified to allow for multiple separator branches on a hemisphere's surface and then trace each branch separately.  This implementation is relatively straightforward and will be the subject of a future study.

We also assume a quasi-steady state with our choice of constant and
uniform solar wind parameters. We look at the magnetic separators late
in simulation time after the magnetopause has achieved a steady
state. \citet{laitinen2007} rotated the IMF by $10^\circ$ every 10
minutes and found that the magnetic separators exhibit a form of
hysteresis.

We have parametrized the magnetic nulls and separators for a small
subset of solar wind conditions, with particular emphasis on IMF clock
angle. A future study should parametrize the magnetic nulls and
separators as a function of solar wind conditions, dipole tilt, and
IMF $B_x$ to develop a predictive capability.

\begin{acknowledgments}
  Support from NSF grant AGS-0953463 (CMK and PAC), NASA grant
  NNX10AN08A (PAC), and NASA West Virginia Space Grant Consortium (CMK) are gratefully acknowledged.  Simulations were
  performed at the Community Coordinated Modeling Center at Goddard
  Space Flight Center through their public Runs on Request system
  (http://ccmc.gsfc.nasa.gov). The CCMC is a multi-agency partnership
  between NASA, AFMC, AFOSR, AFRL, AFWA, NOAA, NSF and ONR. The
  BATS-R-US Model was developed by the Center for Space Environment
  Modeling at the University of Michigan. A large portion of the
  analysis presented here was made possible via the Kameleon and Space
  Weather Explorer software packages provided by the CCMC. The
  Kameleon software has been provided by the Community Coordinated
  Modeling Center at NASA Goddard Space Flight Center
  (http://ccmc.gsfc.nasa.gov) Software Developers: Marlo M. Maddox,
  David H. Berrios, Lutz Rastaetter. The authors would also like to
  thank M. Maddox and D. Berrios for their software support and T. E.
  Moore for interesting discussions. Travel support to the 2013 Geospace Environment Modeling (GEM) Summer Workshop from NSF, CCMC and GEM is gratefully acknowledged (CMK).

\end{acknowledgments}

\bibliographystyle{agufull08}

\end{article}
\end{document}